\documentclass[twocolumn,prl,amsmath,amssymb,floatfix,superscriptaddress,showpacs,longbibliography]{revtex4-1}
\usepackage{graphicx}
\usepackage{dcolumn}
\usepackage{bm}
\usepackage{xspace}
\usepackage{hyperref}
\usepackage{color}
\usepackage{physics}
\usepackage[normalem]{ulem}

\renewcommand{\AA}{\text{\r{A}}}
\def\fgt{Fe$_{3-x}$GeTe$_{2}$\xspace}

\newcommand{\hc}{\text{H.c.}}
\newcommand{\bdsb}[1]{\mathbf{#1}}
\begin{document}


\title{Magnon Damping as a Probe of Kondo Coupling in Magnetically Ordered Systems}
\author{Song~Bao}
\altaffiliation{These authors contributed equally to the work.}
\affiliation{National Laboratory of Solid State Microstructures and Department of Physics, Nanjing University, Nanjing 210093, China}
\affiliation{Collaborative Innovation Center of Advanced Microstructures and Jiangsu Physical Science Research Center, Nanjing University, Nanjing 210093, China}
\author{Yuan~Gao}
\altaffiliation{These authors contributed equally to the work.}
\affiliation{School of Physics, Beihang University, Beijing 100191, China}
\affiliation{Institute of Theoretical Physics, Chinese Academy of Sciences, Beijing 100190, China}
\author{Junsen~Wang}
\altaffiliation{These authors contributed equally to the work.}
\affiliation{Institute of Theoretical Physics, Chinese Academy of Sciences, Beijing 100190, China}
\affiliation{Center of Materials Science and Optoelectronics Engineering, College of Materials Science and Opto-electronic Technology, University of Chinese Academy of Sciences, Beijing 100049, China}
\author{Shin-ichiro~Yano}
\affiliation{National Synchrotron Radiation Research Center, Hsinchu 30077, Taiwan}
\author{Yanyan~Shangguan}
\affiliation{National Laboratory of Solid State Microstructures and Department of Physics, Nanjing University, Nanjing 210093, China}
\author{Zhentao~Huang}
\affiliation{National Laboratory of Solid State Microstructures and Department of Physics, Nanjing University, Nanjing 210093, China}
\affiliation{Institute of High Energy Physics, Chinese Academy of Sciences (CAS), Beijing 100049, China}
\affiliation{Spallation Neutron Source Science Center, Dongguan 523803, China}
\author{Junbo~Liao}
\affiliation{National Laboratory of Solid State Microstructures and Department of Physics, Nanjing University, Nanjing 210093, China}
\author{Wei~Wang}
\affiliation{School of Science, Nanjing University of Posts and Telecommunications, Nanjing 210023, China}
\author{Bo~Zhang}
\author{Shufan~Cheng}
\author{Hao~Xu}
\affiliation{National Laboratory of Solid State Microstructures and Department of Physics, Nanjing University, Nanjing 210093, China}
\author{Zhao-Yang~Dong}
\affiliation{Department of Applied Physics, Nanjing University of Science and Technology, Nanjing 210094, China}
\author{Shun-Li~Yu}
\affiliation{National Laboratory of Solid State Microstructures and Department of Physics, Nanjing University, Nanjing 210093, China}
\affiliation{Collaborative Innovation Center of Advanced Microstructures and Jiangsu Physical Science Research Center, Nanjing University, Nanjing 210093, China}
\author{Wei~Li}
\email{w.li@itp.ac.cn}
\affiliation{Institute of Theoretical Physics, Chinese Academy of Sciences, Beijing 100190, China}
\author{Jian-Xin~Li}
\email{jxli@nju.edu.cn}
\affiliation{National Laboratory of Solid State Microstructures and Department of Physics, Nanjing University, Nanjing 210093, China}
\affiliation{Collaborative Innovation Center of Advanced Microstructures and Jiangsu Physical Science Research Center, Nanjing University, Nanjing 210093, China}
\author{Jinsheng~Wen}
\email{jwen@nju.edu.cn}
\affiliation{National Laboratory of Solid State Microstructures and Department of Physics, Nanjing University, Nanjing 210093, China}
\affiliation{Collaborative Innovation Center of Advanced Microstructures and Jiangsu Physical Science Research Center, Nanjing University, Nanjing 210093, China}

\begin{abstract}
In $d$-electron systems, there can also be intricate interplay between Kondo coupling and magnetic interactions as that in $f$-electron systems, but the underlying mechanism remains elusive. Here, using inelastic neutron scattering, we investigate the temperature evolution of the low-energy spin waves (magnons) in a metallic van der Waals ferromagnet Fe$_{3-x}$GeTe$_{2}$, and observe that the magnon damping diverges at both low and high temperatures and exhibits a minimum at an intermediate temperature. These behaviours are described by a formula that combines logarithmic and power-law terms, representing the dominant contributions from Kondo coupling and thermal fluctuations, respectively. These findings can be explained by considering electron-magnon scattering of spin-flip type within the ferromagnetic Kondo-Heisenberg lattice model, distinct from the original Kondo effect which only considers the coupling between itinerant electrons and isolated impurity spins. These results unveil the intriguing interplay between itinerant electrons and spin waves in metallic 3$d$-electron systems with magnetic order, and provide magnon damping  as a new effective probe of Kondo coupling in metallic quantum magnets, thereby opening new avenues for exploring Kondo physics from the magnon perspective.
\end{abstract}

\maketitle

\begin{figure*}[htb]
\centering
\includegraphics[width=0.9\linewidth]{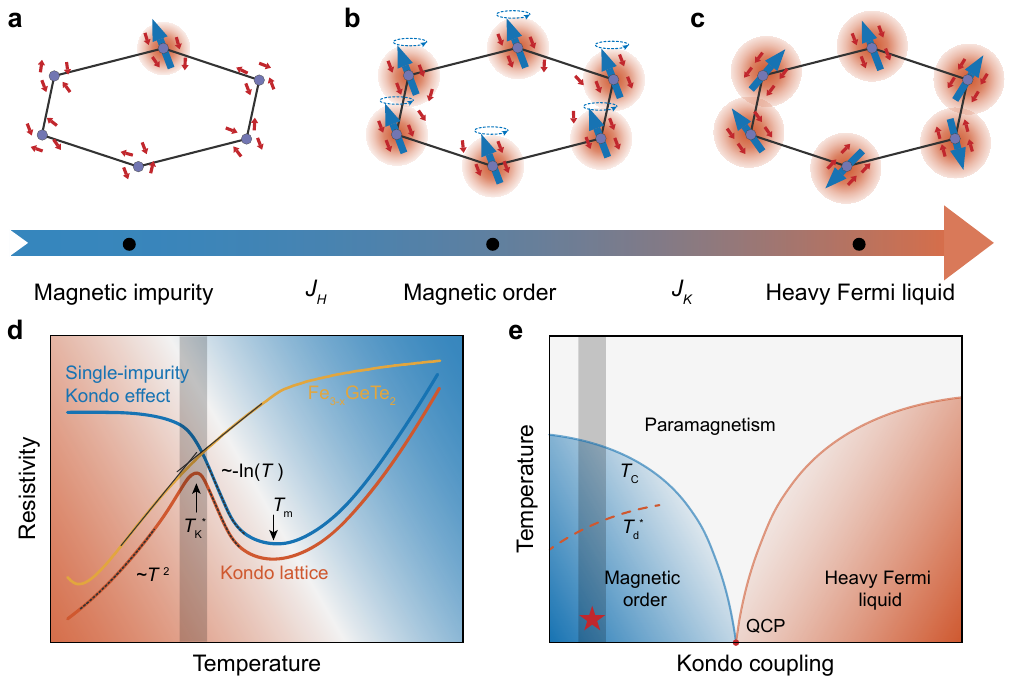}
\caption{\label{fig:sketch}{{\bf Schematic of Kondo screening and resistivity behaviour in different Kondo systems.} {\bf a}-{\bf c}, Schematics illustrating different Kondo screening scenarios based on the concentration of local moments, the strength of ferromagnetic interaction $J_{H}$ and Kondo coupling $J_{K}$. In Kondo-impurity model ({\bf a}), local moments act as magnetic impurities. In ferromagnetic Kondo-Heisenberg lattice model ({\bf b}), representative of the case in \fgt, a dense periodic array of local moments aligns parallel via $J_{H}$, inducing magnetic order. These moments are partially screened by the conduction electron sea through weak $J_{K}$. In the heavy Fermi liquid state ({\bf c}), local moments are fully quenched, resulting in a nonmagnetic heavy-fermion phase driven by strong $J_{K}$. {\bf d}, Temperature dependence of resistivity for the three scenarios depicted in {\bf a}-{\bf c}. In {\bf a} and {\bf c}, incoherent spin-flip scattering between localized spins and conduction electrons results in a resistivity minimum followed by a $\sim -\ln(T)$ divergence below $T_{\rm m}$. At lower temperatures, Kondo singlet formation leads to temperature-independent resistivity in {\bf a}, while in {\bf c}, resistivity follows a Fermi-liquid $T^2$ scaling below the coherence temperature $T_{\rm K}^*$. In \fgt, a slope change in resistivity around 90~K indicates an incoherent-coherent crossover, as represented by the shaded region. {\bf e}, The canonical Doniach phase diagram for a ferromagnetic Kondo lattice as a function of $J_{K}$, showing the magnetically ordered regime and the heavy-fermion regime, separated by a quantum critical point. The shaded region and the star illustrate the phase space to which \fgt belongs. $T_{\rm C}$ is the Curie temperature of the FM order,
and magnon damping reaches its minimum at $T_d^*$.}}
\end{figure*}

The original Kondo effect in dilute magnetic alloys, which takes the resistivity minimum as one of the characteristics, describes the scattering of conduction electrons by magnetic impurities in such systems~\cite{10.1143/PTP.32.37,hewson1997kondo}. This effect is classified as the single-impurity Kondo problem and is now well understood based on the spin-flip scatterings of conduction electrons off the single-impurity spins~(Fig.~\ref{fig:sketch}a)~\cite{10.1143/PTP.32.37,hewson1997kondo}. The concept of Kondo physics has been extended to include the correlated electron systems, specifically those consisting of a dense periodic array of local moments interacting with the conduction electron sea through an antiferromagnetic Kondo interaction, referred to as a Kondo lattice~(Fig.~\ref{fig:sketch}b-d)~\cite{coleman2015introduction,pavarini2015many,hewson1997kondo,RevModPhys.56.755}. The Kondo-lattice model was initially proposed to describe the heavy-fermion state in $f$-electron systems, particularly in intermetallics containing Ce, Yb, or U elements~\cite{coleman2015introduction,pavarini2015many,hewson1997kondo,RevModPhys.56.755,RevModPhys.92.011002,RevModPhys.79.1015}. The competition between the Kondo effect and couplings of localized spins underlies various intriguing phenomena in these heavy-fermion compounds (Fig.~\ref{fig:sketch}e), including quantum criticality~\cite{schroder2000onset,PhysRevLett.89.056402,custers2003break,shen2020strange}, strange-metal behaviour~\cite{shen2020strange,RevModPhys.73.797}, magnetism~\cite{PhysRevLett.123.106402,doi:10.1126/sciadv.aaw9061,PhysRevB.76.125101,PhysRevLett.121.057201} and superconductivity\cite{Aoki2001,RevModPhys.81.1551}, which can be tuned by chemical doping, magnetic field or pressure~\cite{schroder2000onset,shen2020strange,PhysRevLett.89.056402,custers2003break,RevModPhys.73.797}.
In the strong Kondo coupling regime, the hybridization between localized $f$ electrons and conduction electrons leads to the quenching of local moments and the expansion of the Fermi surface, resulting in the formation of a heavy Fermi-liquid state~(Fig.~\ref{fig:sketch}c,d)~\cite{coleman2015introduction,pavarini2015many,hewson1997kondo,RevModPhys.56.755,RevModPhys.92.011002,RevModPhys.79.1015,DONIACH1977231,yang2008scaling}. Magnetic order can arise when the strength of the effective magnetic interaction between localized spins exceeds the Kondo coupling. Moreover, the coexistence of magnetic order with Kondo effect~\cite{PhysRevLett.123.106402,doi:10.1126/sciadv.aaw9061,PhysRevB.76.125101,PhysRevLett.121.057201} or superconductivity~\cite{Aoki2001,RevModPhys.81.1551} can be observed, particularly in systems where the duality of $f$ electrons emerges due to the mixed-valence situation and the multiorbital nature~\cite{PhysRevLett.123.106402,doi:10.1126/sciadv.aaw9061,PhysRevB.76.125101,PhysRevLett.121.057201,Aoki2001,RevModPhys.81.1551}.

\begin{figure*}[t]
	\centering
	\includegraphics[width=0.9\linewidth]{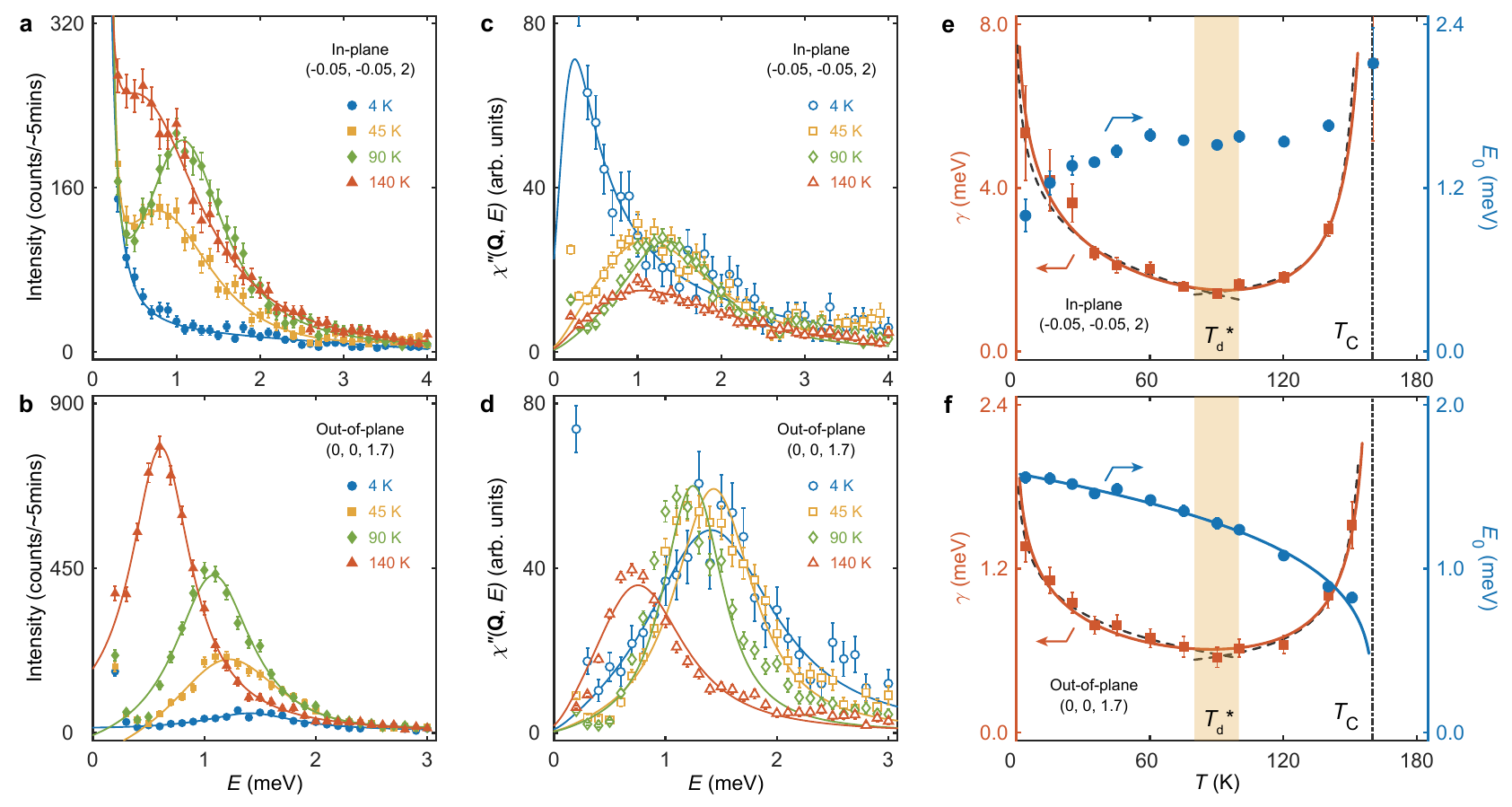}
	\caption{\label{fig:escans}{{\bf Nonmonotonic temperature evolution of magnon damping.} {\bf a},{\bf b}, Temperature evolution of the energy scans off the zone center at the in-plane position (-0.05,\,-0.05,\,2) and out-of-plane position (0,\,0,\,1.7), respectively. Different symbols represent different temperatures. Lines are guides to the eye with Lorentzian fittings. {\bf c},{\bf d}, The corresponding data from {\bf a} and {\bf b}, corrected by the Bose population factor. Solid lines are fits using the DHO formula convoluted with the instrumental resolution, as discussed in the text. {\bf e},{\bf f}, Extracted damping rate $\gamma$ (red squares, left axis) and magnon energy $E_0$ (blue circles, right axis) as a function of temperature. The dashed curves are guides to the eye, fitted by either a logarithmic or power-law function. The red solid curves are fits using a linear combination of these two terms. The blue solid curve in {\bf f} is a fit with a simple power-law function. Throughout the paper, errors represent one standard deviation.}}
\end{figure*}

More recently, the unexpected heavy-fermion state has also been observed in certain $d$-electron transition metals, which reside in the intermediate regime between itineracy and localization of $d$ electrons~\cite{PhysRevLett.78.3729,PhysRevLett.85.1052,doi:10.1143/JPSJ.73.2373,PhysRevLett.111.176403,PhysRevLett.116.147001,2011arXiv1103.5073Z,PhysRevLett.124.087202,Kim2022,Zhangeaao6791,doi:10.1021/acs.nanolett.1c01661,PhysRevX.12.011022}. Understanding magnetism within the framework of this duality and the applicability of the Kondo-lattice model to such systems are currently active topics of research~\cite{PhysRevLett.78.3729,doi:10.1143/JPSJ.73.2373,PhysRevLett.116.147001,2011arXiv1103.5073Z,PhysRevLett.124.087202,Kim2022,Zhangeaao6791,doi:10.1021/acs.nanolett.1c01661,PhysRevX.12.011022}.
The van der Waals (vdW) metallic ferromagnet Fe$_3$GeTe$_2$ is one such example material~\cite{Deng2018,fei2018two,PhysRevB.93.144404,PhysRevB.102.161109,Zhangeaao6791,doi:10.1021/acs.nanolett.1c01661,PhysRevX.12.011022}.
Due to the coexistence of localized and itinerant 3$d$ electrons, there are conflicting reports regarding the microscopic origin of the magnetism in \fgt~\cite{Zhangeaao6791,doi:10.1021/acs.nanolett.1c01661,PhysRevX.12.011022,Deng2018,doi:10.7566/JPSJ.82.124711,PhysRevB.101.201104,PhysRevB.99.094423,PhysRevB.106.L180409}. In our earlier work, we have reconciled the debate by showing that the ferromagnetism has a dual origin, with local moments and itinerant electrons contributing to the low-energy spin waves and columnlike continua, respectively~\cite{PhysRevX.12.011022}. Moreover, there is accumulating evidence for the Kondo coupling between these two components, such as the Fano resonance feature~\cite{Zhangeaao6791,doi:10.1021/acs.nanolett.1c01661}, large effective electron mass~\cite{Zhangeaao6791,PhysRevB.101.201104,PhysRevB.93.144404}, and the incoherent-coherent crossover in transport and magnetic measurements~\cite{Zhangeaao6791}. In particular, the resistivity curve of \fgt reveals a slope change at a characteristic temperature of approximately 90~K, as depicted in Fig.~\ref{fig:sketch}d. Below this temperature, the Kondo-lattice behaviour emerges in the magnetic ordering phase, accompanied by the enhancement of Fermi surface volume and effective electron mass~\cite{Zhangeaao6791}.
Notably, this resistivity behaviour aligns with observations in other $d$-electron systems exhibiting a heavy-fermion state~\cite{PhysRevLett.116.147001,PhysRevLett.111.176403,PhysRevLett.85.1052}, but is distinct from either the Kondo-impurity model or the Kondo lattice in $f$-electron systems~(Fig.~\ref{fig:sketch}d). Since our earlier work has indicated that Kondo screening significantly enhances spin-wave damping at 4~K compared to 100~K~\cite{PhysRevX.12.011022}, a question naturally arises: will the damping follow certain scaling behaviour due to the Kondo coupling effect?

\begin{figure*}[t]
\centering
\includegraphics[width=0.9\linewidth]{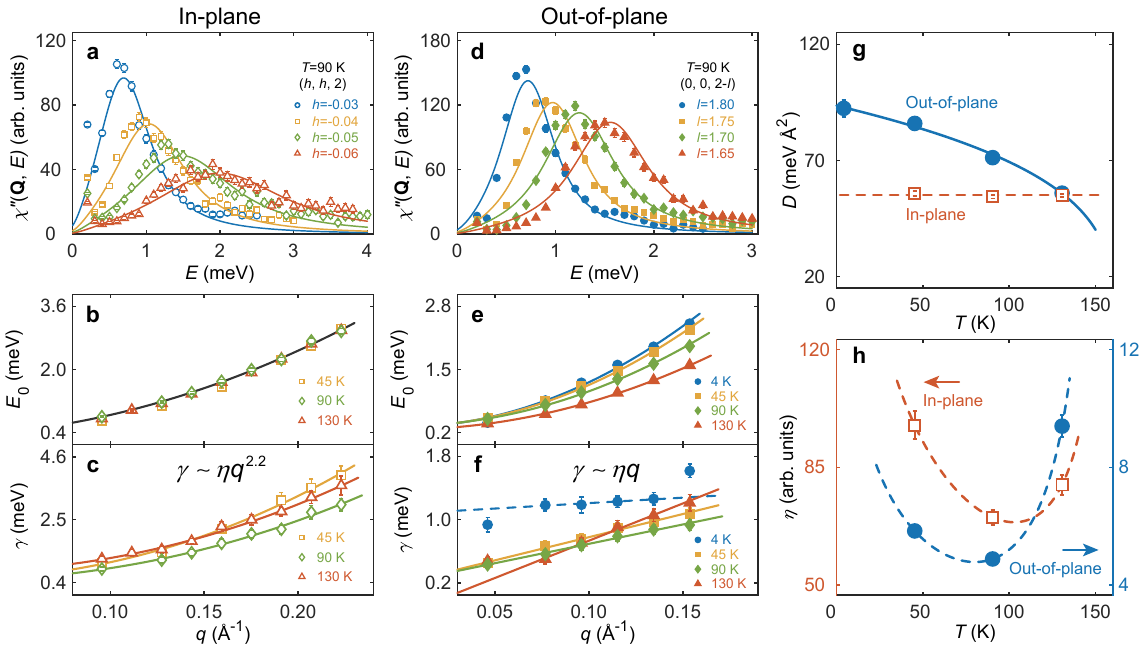}
\caption{\label{fig:qdependence}{{\bf Momentum dependence of magnon damping.} {\bf a}, Bose-factor-corrected energy scans at different ${\bf Q}$s at 90~K along the in-plane direction ($h$,\,$h$,\,2). Solid lines are fits using the DHO formula convoluted with the instrumental resolution. {\bf b},{\bf c}, Extracted $E_0$ and $\gamma$ for the in-plane direction as a function of the reduced wave vector $q$ at 90~K and other temperatures. {\bf d}-{\bf f}, Same as in {\bf a}-{\bf c} but for the out-of-plane direction (0,\,0,\,2-$l$). Lines in {\bf b} and {\bf e} are fits using quadratic functions. Lines in {\bf c} and {\bf f} are obtained with quasi-quadratic and linear fittings, respectively. {\bf g},{\bf h}, In-plane and out-of-plane spin-wave stiffness and damping coefficient as a function of temperature. The solid curve in {\bf g} is a fit with a simple power-law function and the dashed line is a guide to the eye. Dashed curves in {\bf h} are guides to the eye with the form of Eq.~\ref{InKondo}. $D$ denotes the spin-wave stiffness, and $\eta$ is the coefficient of damping indicated in \textbf{c,f}.}}
\end{figure*}

Here we report inelastic neutron scattering (INS) measurements on the temperature evolution of low-energy spin waves in Fe-deficient \fgt single crystals with Curie temperature $T_{\rm C}\sim160$~K. Upon cooling from $T_{\rm C}$, we observe that the magnon damping rate initially decreases, reaching a minimum around $T_{\rm d}^*\sim90$~K, before rapidly increasing again. This behaviour follows a scaling relation combining both logarithmic and power-law terms. Additionally, we find that the in-plane Kondo coupling is stronger than the out-of-plane coupling, resulting in a softening of in-plane magnons at low temperatures. Using the state-of-the-art tensor-network algorithm~\cite{SETTN,Chen2018,tanTRG2023,TDVP2011,TDVP2016,MPSManifold2014}, we reproduce the observed magnon damping minimum and logarithmic scaling within a ferromagnetic Kondo-Heisenberg (FMKH) model. Although the form is analogous to the original Kondo effect describing the coupling between itinerant electrons and impurity spins~\cite{10.1143/PTP.32.37,hewson1997kondo}, the underlying mechanism is essentially different --- here, itinerant $d$ electrons scatter the collective excitations of local moments via a spin-flip process induced by Kondo coupling. Our results provide compelling evidence for the existence of Kondo effect in the metallic ferromagnet \fgt, and demonstrate magnon damping as an effective probe for the Kondo coupling in systems where magnetic order and Kondo effect coexist.

\medskip
\noindent{\bf Results}\\
\noindent{\bf Damping rate minimum and logarithmic scaling}\\
Figure~\ref{fig:escans}a,b shows the energy scans at various temperatures for the off-centered in-plane and out-of-plane positions, respectively. As no sizable spin gap was observed at the zone center (0,\,0,\,2)~\cite{PhysRevX.12.011022}, we chose $\bf{Q}$ positions away from the zone center in order to track the temperature evolution of the inelastic magnon peak. Intriguingly, for (-0.05,\,-0.05,\,2) within the in-plane direction (Fig.~\ref{fig:escans}a), no magnon peak is observed at 4~K. Instead, the magnon peak emerges at higher temperatures, with its center exhibiting a nonmonotonic change approaching $T_{\rm C}$. For (0,\,0,\,1.7) in the out-of-plane direction (Fig.~\ref{fig:escans}b), the peak center consistently shifts to lower energies with increasing temperature. To eliminate the influence of the Bose statistics, we correct the raw data with the Bose population factor, and plot them in Fig.~\ref{fig:escans}c,d. We fit these corrected results by convoluting the momentum- and energy-dependent instrumental resolution with a damped harmonic oscillator (DHO) formula, which is applicable to damped spin waves~\cite{zhao2009spin,nc11_3076,PhysRevX.12.011022,andreas2002}. The DHO formula has a form of $\chi''({\bf Q},E)\propto\gamma{E_0}E/[(E^2-{E_0}^2)^2+(\gamma E)^2]$, where $E_0$ is the magnon energy and $\gamma$ is the damping rate (the inverse of $\gamma$ is proportional to the lifetime of the damped magnons)~\cite{zhao2009spin,nc11_3076,PhysRevX.12.011022,andreas2002}. Based on these DHO fittings in Fig.~\ref{fig:escans}c,d, we can extract the intrinsic damping rate and magnon energy. We emphasize that $E_0$ represents the fitted magnon energy, which does not necessarily coincide with the peak maximum in the measured spectra. This is particularly the case for strongly damped magnons; for example, at 4~K, the peak maximum appears at a much lower energy than $E_0$ (Fig.~\ref{fig:escans}c).

Figure~\ref{fig:escans}e,f presents the extracted $\gamma$ and $E_0$ plotted as a function of temperature, showcasing the most intriguing results of this work. It is found that for both in-plane and out-of-plane directions, $\gamma$ shows an upturn toward both 0~K and $T_{\rm C}$, causing a minimum around $T_{\rm d}^*\sim 90$~K. While disorder or vacancies may affect local moments and make spin waves broader, such effect, if present, is more likely to be temperature independent, and therefore would not be responsible for the observed nonmonotonic temperature dependence. On the other hand, this phenomenon is reminiscent of the resistivity minimum caused by the Kondo effect in the original single-impurity Kondo model~(Fig.~\ref{fig:sketch}d)~\cite{coleman2015introduction,pavarini2015many,10.1143/PTP.32.37,hewson1997kondo}, where thermodynamic and transport properties depend logarithmically on temperature as $-\ln (T)$~\cite{coleman2015introduction,pavarini2015many,10.1143/PTP.32.37,hewson1997kondo}. Inspired by this, we use a similar logarithmic term to describe the divergent behaviour toward 0~K. In the meantime, a power-law term describing the spin-wave damping in the hydrodynamic regime is also required to explain the divergence toward $T_{\rm C}$~\cite{PhysRevB.14.4923,PhysRev.177.952}. Actually, these two effects should exist simultaneously over the entire temperature range below $T_{\rm C}$. Therefore, the general formula should consist of the linear combination of the two terms, which reads as,
\begin{equation}\label{InKondo}
 \gamma(T)=A{\rm In}(T_{\rm d}^*/T)+B(1-T/T_{\rm C})^{-\nu}.
\end{equation}
To fit the data, we fix $T_{\rm d}^*=90$~K and $T_{\rm C}=160$~K, and allow $A$,\,$B$,\,$\nu$ to be free.
The fitting results well reproduce the unusual nonmonotonic temperature evolution of $\gamma$ for both in-plane~(Fig.~\ref{fig:escans}e) and out-of-plane~(Fig.~\ref{fig:escans}f) directions. Specifically, we obtain $A_{\rm in}=1.617\pm0.204$, $B_{\rm in}=0.834\pm0.100$, and $\nu_{\rm in}=0.715\pm0.087$ for the in-plane direction, and $A_{\rm out}=0.356\pm0.024$, $B_{\rm out}=0.406\pm0.025$, and $\nu_{\rm out}=0.490\pm0.040$ for the out-of-plane direction. The distinct parameter values for the in-plane and out-of-plane directions indicate the presence of magnetic anisotropy in this material.

The extracted magnon energy $E_0$ shown in Fig.~\ref{fig:escans}e,f exhibits distinct behaviours for the in- and out-of-plane directions.
With increasing temperature, it is found that the out-of-plane $E_0$ softens through a power law while approaching $T_{\rm C}$~(Fig.~\ref{fig:escans}f), a behaviour similar to the temperature evolution of the magnetization in \fgt~\cite{doi:10.7566/JPSJ.82.124711,PhysRevB.96.144429}.
On the other hand, the in-plane $E_0$ keeps almost constant across $T_{\rm d}^*$ and slightly softens upon cooling to lower temperatures~(Fig.~\ref{fig:escans}e). We attribute these two distinct behaviours of $E_0$ to different degrees of Kondo coupling along the two directions. Notably, the magnitude of $\gamma$ for the in-plane direction is much larger than that for the out-of-plane direction, being approximately four times greater at 4~K. This anisotropy can be explained by the quasi-two-dimensional vdW structure of \fgt, which promotes a more itinerant electron character in the $a$-$b$ plane and results in stronger in-plane Kondo coupling. Consequently, the in-plane Kondo screening markedly reduces the intralayer exchange coupling between local moments, leading to the observed decrease in the in-plane $E_0$ at low temperatures (Fig.~\ref{fig:escans}e).

\begin{figure}[t]
\centering
\includegraphics[width=1.0\linewidth]{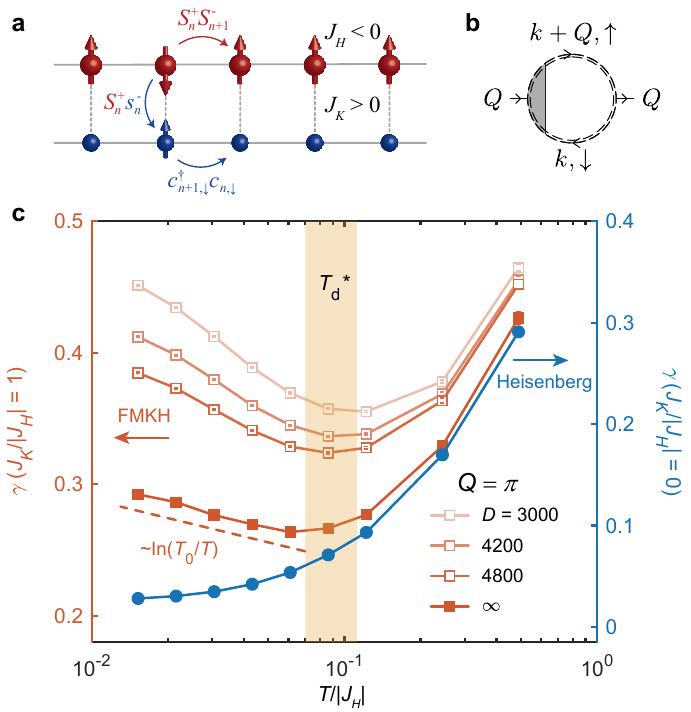}
\caption{\label{fig:FMKH}{{\bf Damping rate in the FMKH chain obtained from tensor-network calculations.} {\bf a}, Schematic illustration of the FMKH model, where the spins are coupled via Heisenberg interaction~($J_H<0$), and the spin flip can dissipate into the itinerant electron host through the Kondo coupling $J_K$. {\bf b}, Feynman diagram of magnon self-energy due to electron-magnon scattering with the renormalized vertex (the gray area) responsible for the anomalous decay of magnon, where the double-dashed lines represent full electron propagator and the thick solid line is the magnon propagator. {\bf c}, Calculated damping rate for FMKH model in the FM phase with ${J}_{K}/|J_{H}|$=1, as compared to that in the pure Heisenberg model. Results for the FMKH model are obtained with different retained bond dimensions $D$, while the results for the pure Heisenberg model are obtained by exact diagonalization.}}
\end{figure}

\medskip
\noindent{\bf Momentum dependence of the scaling behaviour}\\
To examine whether these behaviours can also be observed in other momenta, we plot the results of the low-energy spin-wave excitations and their damping rate as a function of momentum in Fig.~\ref{fig:qdependence}.
We focus on the excitations at 90~K, where the spin waves are most coherent with the least damping. The energy scans along [110] and [001] directions are plotted in Fig.~\ref{fig:qdependence}a,d. As ${\bf Q}$ increases, the peak centers shift gradually to higher energies, accompanied by a broadening of the linewidth and a weakening of the scattering intensities, indicating the nature of damped spin waves. We perform DHO fittings to these scans to extract the momentum dependence of $E_0$~(Fig.~\ref{fig:qdependence}b,e) and $\gamma$~(Fig.~\ref{fig:qdependence}c,f) for the two directions. The dispersions are fitted using the quadratic relation $E_0=\Delta+D{\bf Q}^2$, where $\Delta$ and $D$ are the spin gap and spin-wave stiffness, respectively. The fittings yield $D$ of $54.3\pm0.8~{\text{meV~{\AA}}}^2$ for the [110] direction~(Fig.~\ref{fig:qdependence}b) and $71.2\pm1.3~{\rm {meV~{\AA}}}^2$ for the [001] direction~(Fig.~\ref{fig:qdependence}e) at 90~K. For $\gamma$, we found a quasi-quadratic dependence ($\sim q^{2.2}$) in the in-plane direction (Fig.~\ref{fig:qdependence}c) and a linear dependence in the out-of-plane direction (Fig.~\ref{fig:qdependence}f).

The low-energy spin waves at other temperatures like 4, 45 and 130~K are also measured. With increasing temperature, we find that the spin-wave dispersion remains nearly unchanged for the in-plane direction but softens considerably for the out-of-plane direction~(Fig.~\ref{fig:qdependence}b,e). Note that since the in-plane spin waves are heavily damped at 4~K, it is challenging to extract a reasonable dispersion. We find that the original $q$-dependences of $\gamma$ at 90~K remain applicable at other temperatures, {\it i.e.}, $\gamma_{\rm in}\sim\eta q^{2.2}$ and $\gamma_{\rm out}\sim\eta q$, although the magnitude and coefficient $\eta$ of the damping vary due to the temperature dependence of the Kondo screening effect~(Fig.~\ref{fig:qdependence}c,f).
To further illustrate the temperature evolution of the spin waves, we plot the extracted spin-wave stiffness $D$ and damping coefficient $\eta$ as a function of temperature in Fig.~\ref{fig:qdependence}g,h. The out-of-plane $D$ follows a typical power-law relation with temperature, while the in-plane one remains nearly constant. For $\eta$, it follows a nonmonotonic dependence and can still be elucidated using Eq.~\ref{InKondo} with fixed $T_{\rm d}^*=90$~K and $T_{\rm C}=160$~K for both directions. Together with Fig.~\ref{fig:escans}e,f, these findings uncover a universal law independent of the momentum governed by the Kondo physics in \fgt.


\medskip
\noindent{\bf FMKH model and spin-flip scattering mechanism}\\
Notably, the temperature at which this damping minimum occurs (Fig.~\ref{fig:escans}e,f and Fig.~\ref{fig:qdependence}h) is close to the incoherent-coherent crossover observed in the resistivity curve (Fig.~\ref{fig:sketch}d). Below the intermediate temperature, the localized electronic state hybridizes with the itinerant electronic state, enlarging the Fermi surface volume and effective electron mass~\cite{Zhangeaao6791}. Consequently, from the perspective of conduction electrons, the slope of resistivity changes~(Fig.~\ref{fig:sketch}d), consistent with other experimental observations~\cite{Zhangeaao6791,doi:10.1021/acs.nanolett.1c01661,PhysRevB.93.144404,PhysRevX.12.011022}, manifesting heavy-fermion behaviour. On the other hand, from the perspective of spin waves, these processes provide additional decay channels for magnons, resulting in a significant damping of magnons at low temperatures.

To understand the anomalous magnon damping behaviour microscopically, we develop an FMKH model, considering the dominant role of direct ferromagnetic (FM) exchange interactions between localized $d$ moments and their competition with Kondo coupling in \fgt (Methods). The strong direct exchange interactions arise from the spatially extended nature of the outermost 3$d$ orbitals, whose significant overlap stabilizes robust long-range FM order and a relatively high $T_{\rm C}$. This stands in sharp contrast to conventional $f$-electron Kondo-lattice systems, where the highly localized $f$ orbitals are effectively screened by conduction electrons and the local $f$ moments couple primarily via weaker, indirect Ruderman-Kittel-Kasuya-Yosida (RKKY) interactions, which often favor antiferromagnetic order and yield lower ordering temperatures~\cite{schroder2000onset,PhysRevLett.89.056402,custers2003break,PhysRevLett.123.106402,doi:10.1126/sciadv.aaw9061}. In \fgt, the localized $d$ moments are only partially screened by itinerant electrons~(Fig.~\ref{fig:sketch}b). Within this framework, magnons as collective modes of local moments are interacting with conduction electrons via Kondo coupling~\cite{Deng2018,Tang2023}, giving rise to the Kondo effect~\cite{Zhangeaao6791, doi:10.1021/acs.nanolett.1c01661, PhysRevX.12.011022}. To characterize such effect, we employ tensor-network methods to compute the dynamical properties of a minimal 1D FMKH system at finite temperature, represented by a two-leg ladder: one leg as a chain of local moments and the other as itinerant electrons~(Fig.~\ref{fig:FMKH}a) (see Methods for details). The calculated damping rate for the FM phase of the FMKH model (${J}_{K}/|J_{H}|$=1) is shown in Fig.~\ref{fig:FMKH}c. A clear damping minimum at $T_{\rm d}^*$ is observed, contrasting with the monotonic behaviour in the pure Heisenberg model (${J}_{K}/|J_{H}|$=0). Notably, our simulations also reveal a logarithmic scaling $\gamma(T)\sim{\rm ln}(T_{\rm 0}/T)$ below $T_{\rm d}^*$. This damping minimum and logarithmic scaling behaviour can be observed in various momenta and for different Kondo couplings in the FM phase (Supplementary Fig.~2), indicating that it is a universal feature arising from the scattering between magnons and itinerant electrons. 

We emphasize that the key mechanism underlying the anomalous magnon decay is spin-flip scattering between conduction electrons and magnons induced by the Kondo coupling. In these processes, a magnon decays into a fermionic particle-hole pair, providing an additional channel for damping~(Fig.~\ref{fig:FMKH}a). In the weak-coupling limit ($J_{K}\ll |J_{H}|$) relevant to \fgt (Fig.~\ref{fig:sketch}e), the magnon self-energy is given diagrammatically in Fig.~\ref{fig:FMKH}(b), where the renormalized vertex plays a critical role in enabling anomalous magnon decay~\cite{Fetter2012}.
In sharp contrast to the usual vertex correction due to electron-phonon coupling --- which is negligible due to Migdal theorem~\cite{Migdal1958} --- here the vertex correction of $J_K$ in Fig.~\ref{fig:FMKH}b is crucial at low temperatures. By applying the renormalization of the Kondo coupling, formally similar to the single-impurity Kondo problem but here used to describe the momentum-dependent renormalization and damping of magnons, a ${\rm log}(1/T)$-like temperature dependence of the magnon self-energy can be obtained. Furthermore, as the kinematically allowed phase space for magnon-electron scattering increases with increasing magnon momentum $\bf Q$ (See Supplementary Fig.~4), the magnon damping becomes stronger with ${\bf Q}$ at low temperatures, in agreement with the experimental observations in Fig.~\ref{fig:qdependence}.

\medskip
\noindent{\bf Discussion}\\
Our study here uncovers a magnon damping minimum with logarithmic scaling at low temperatures and an increase in damping with momentum in \fgt. Logarithmic scaling as a function of temperature is a hallmark of the Kondo coupling, previously manifested in the thermodynamic and transport measurements~\cite{10.1143/PTP.32.37, hewson1997kondo, coleman2015introduction, pavarini2015many, shen2020strange}, as well as in the evolution of electronic structures near the Fermi level~\cite{doi:10.1073/pnas.2001778117, Poelchen2020, PhysRevB.107.L201104}, and now extended to magnon damping rates. It is worth noting that a nonmonotonic temperature dependence of spin relaxation rates, sharing a similar form, have been reported in certain Ce-based heavy-fermion compounds~\cite{Knopp1989, KNOPP1988341, 10.1063/1.335137, PhysRevLett.54.230}. However, those studies addressed momentum-independent single 4$f$-moment fluctuations in paramagnetic or weakly ordered states, probed by quasielastic neutron scattering or nuclear magnetic resonance. The extracted relaxation rates of local moment were interpreted within the framework of the single-impurity Kondo model~\cite{Knopp1989, KNOPP1988341, 10.1063/1.335137, PhysRevLett.54.230}.

In sharp contrast, our work presents, to the best of our knowledge, the first momentum-resolved INS measurement of a magnon damping phenomenon for the collective excitations of local moments (as opposed to single spin) in a magnetically ordered 3$d$-electron Kondo lattice. Here the damping rates reflects the lifetime of magnons, highlighting more complex many-body effects that necessarily involve both the collective spin excitations and itinerant electrons. Building upon this distinction, our results extend the logarithmic scaling for the damping rates of collective spin excitations. The Kondo coupling, while exhibiting moderate strength and consequently giving rise to a moderate Kondo effect in their electronic properties, strongly influences the magnetic excitations of local moments in \fgt. Importantly, this work demonstrates that magnon damping can serve as a very sensitive indicator of the Kondo coupling. Our findings pave the way for exploring Kondo physics, particularly that related to spin dynamics, within metallic quantum ferromagnets and other spin ordered systems.

As we demonstrate in Fig.~\ref{fig:FMKH}c, experimental observations can be successfully reproduced with a 1D FMKH chain using tensor-network methods. Extending tensor-network calculations of spin dynamics to the 2D FMKH model remains challenging, with convergence issues in the real-time evolution of the dynamics calculations yet to be resolved~\cite{Gao2023KHM}. Nevertheless, field-theoretical analysis shows that the same scenario may also hold independent of dimensionality~\cite{Gao2023KHM}. Therefore, we believe the 1D FMKH model serves as a minimal model that captures the essential damping mechanism of a FM Kondo lattice---the competition between FM exchange and Kondo coupling. It is worth mentioning that enhancing the Kondo coupling by applying pressure on \fgt has been proposed as a means to suppress the Curie temperature $T_{\rm C}$ by approaching a quantum critical point~\cite{Gao2023KHM}, as illustrated in Fig.~\ref{fig:sketch}e. Future theoretical and experimental studies, particularly high-pressure experiments, will be essential to explore these possibilities. Finally, an interesting open question raised by this study is whether logarithmic scaling and the minimum in magnon damping is a universal feature of Kondo systems or not. Systematically investigating the role of Kondo coupling in the spin-wave excitations within the magnetically ordered regime of the Doniach-type phase diagram in both $d$- and $f$-electron systems~\cite{PhysRevLett.123.106402,doi:10.1126/sciadv.aaw9061, PhysRevB.76.125101, PhysRevLett.121.057201, Aoki2001, RevModPhys.81.1551, PhysRevLett.78.3729, PhysRevLett.85.1052, doi:10.1143/JPSJ.73.2373, PhysRevLett.111.176403, PhysRevLett.116.147001, 2011arXiv1103.5073Z, PhysRevLett.124.087202, Kim2022, Zhangeaao6791, doi:10.1021/acs.nanolett.1c01661, PhysRevX.12.011022}, should help elucidate the universality of this phenomenon.


\bigskip
\noindent {\bf Methods}\\
\noindent{\bf Crystal growth and INS experiments.}
High-quality single crystals of \fgt were grown using self-flux method, as described in Refs.~\onlinecite{PhysRevX.12.011022,PhysRevB.93.014411}. This growth method naturally introduces Fe vacancies, which reduce both the Curie temperature and the saturation magnetization. To make sure that different pieces of single crystals had the same amount of Fe vacancies, we measured the magnetization on four randomly picked single crystals from different growth batches. As shown in Supplementary Fig.~5, they have nearly identical susceptibility curves with a single $T_{\rm C}\sim160$~K, regardless the growth batch. This is in good agreement with our prior magnetization and resistivity measurements, as well as with the temperature evolution of the magnetic Bragg peak~(Ref.~\onlinecite{PhysRevX.12.011022}). These results confirm the homogeneity of the crystals used in this study. The INS experiments were performed on Sika, a cold-neutron triple-axis spectrometer located at the OPAL facility of ANSTO in Australia~\cite{Wu_2016,Yano2020}. The \fgt sample array used in the measurements was the same as that employed in Ref.~\onlinecite{PhysRevX.12.011022}. It was composed of well coaligned single crystals with a mosaic of $1.6^\circ$ within the $a$-$b$ plane and $1.7^\circ$ along the $c$ axis (Supplementary Fig.~5), and mounted in the $(H,\,H,\,L)$ scattering plane. Measurements were carried out using a fixed final-energy mode with $E_{\rm f}=5.0$~meV, where both incident and final neutron energies were determined by pyrolytic graphite (002) crystals. Double and vertical focusing modes were applied for the monochromator and analyzer, respectively. A cooled Be filter was placed after the sample during the first experimental run (corresponding to Fig.~2) and removed in the subsequent run, resulting in higher detected neutron intensity (corresponding to Fig.~3). An open-open-$60'$-$60'$ collimation was used to strike a fine balance between neutron flux and experimental resolution. These settings gave an energy resolution of $\sim0.2$ meV at the elastic line. To ensure sufficient statistics, each inelastic data point was counted for $\sim5$ minutes. Throughout the paper, the wavevector ${\bf Q}$ is expressed as $(H,\,K,\,L)$ in reciprocal lattice unit (r.l.u.) of $(a^*,\,b^*,\,c^*)=(4\pi/\sqrt{3}a,\,4\pi/\sqrt{3}b,\,2\pi/c)$ with refined lattice parameters $a=b=3.946~{\rm \AA}$ and $c=16.357~{\rm \AA}$ in a hexagonal structure.

\smallskip
\noindent{\bf FMKH model and tensor-network calculations.}
The FMKH model considered in this work reads
\begin{equation}
\begin{split}
H =  &- t  \sum_{n, \sigma} \left(c_{n+1,\sigma}^\dagger c_{n,\sigma}
+ \hc\right) \\&+ J_{H} \sum_{n} {\bdsb S}_n \cdot {\bdsb S}_{n+1} +
J_{K} \sum_{n} {\bdsb S}_n \cdot {\bdsb s}_n,
\label{Eq:Ham}
\end{split}
\end{equation}
where $c^\dagger_{n,\sigma}$ ($c_{n,\sigma}$) is the creation
(annihilation) operator of the conduction electron at site $n$
with spin $\sigma \in \{\uparrow, \downarrow\}$, and $\bdsb S_n$
($\bdsb s_n$) represents the spin operator of local moment
(conduction electron). The finite-temperature dynamical properties of
this model are calculated by the tensor-network methods
(for 1D chain)~\cite{SETTN,Chen2018,tanTRG2023,TDVP2011,
TDVP2016,MPSManifold2014}, and analyzed by perturbative
calculations (also for higher dimensions~\cite{Gao2023KHM}). The tensor-network calculation process
can be divided into two parts: first, obtaining the equilibrium thermal
density matrix, and second, performing real-time evolution on that density
matrix.  The equilibrium thermal density matrix and related thermodynamic
properties can be obtained by the matrix product operator (MPO)
based tensor-network methods~\cite{SETTN,Chen2018,
tanTRG2023}. Given the MPO representation of the thermal density matrix,
we generalize the tangent-space approach for real-time evolution
\cite{TDVP2011,TDVP2016,MPSManifold2014} to the mixed states
and study the temperature evolution of the spectral density.

In the calculations, FM coupling strength $J_H=-1$ is taken as the energy unit and the hopping energy is set as $t/|J_H|= 2$ throughout the calculations. The simulations are restricted within the 2-leg ladder structure of size $N=L\times2$ with the length $L=16$. This moderate system size is chosen so as to guarantee converged dynamical data at finite temperature. One chain comprises local moments, while the other represents itinerant electrons. These two chains are coupled through Kondo interactions. The retained bond dimension is $D = 1600$, which guarantees accurate results till low temperature $T/|J_H| \simeq 0.01$. Given the MPO representation of the thermal density matrix $\rho(\beta)= e^{-\beta H}$ ($\beta\equiv 1/T$), the dynamical properties at temperature $T$ can be simulated through a successive real-time evolution. The spectral function $\chi''(Q,E) \approx -\int_0^{t_{\rm max}} dt~{\rm Im}[g(Q,t)] \sin{(E t)} \cdot W({t}/{t_{\rm max}})$ is of central interest in studying the dynamical properties at finite temperature, where $W$ is the Hanning window function, $t_{\rm max}$ is the evolution time, and the (dynamical) correlation function is $g(Q,t) = \langle A_Q(t) A_Q^\dagger \rangle_\beta$, with $A_Q = \frac{1}{\sqrt{L}} \sum_n^L S_n^+ e^{{\rm i} Q n}$ being the Fourier transformed spin operator. The time-dependent correlation function $g(Q,t)$ can be obtained by tangent-space approach~\cite{TDVP2011,TDVP2016,MPSManifold2014}. In the real-time evolution calculations, we consider a $16\times2$ lattice and evolution time up to $t_{\rm max}= 20/|J_H|$, with $D = 4800$ bond states retained. The damping rate $\gamma$ is obtained through the DHO fitting of numerical results (see more details in Supplementary Information).

\bigskip
\noindent {\bf Data availability}\\
\noindent Data supporting the findings of this study are available from the corresponding author J.W. upon reasonable request.\\

\bigskip
\noindent {\bf Code availability}\\
\noindent The codes used for the tensor-network calculations in this study are available from the corresponding author J.W. upon reasonable request.\\

\bigskip
\noindent {\bf Acknowledgements}\\
\noindent The work was supported by the National Natural Science Foundation of China with Grant No.~12225407 (J.W.), the National Key Projects for Research and Development of China with Grant Nos.~2021YFA1400400 (J.W. and J.-X.L.) and 2024YFA1409200 (J.W., S.B., and W.L.), the National Natural Science Foundation of China with Grant Nos.~12434005 (J.W., J.-X.L., S.-L.Y., and Y.S.), 12404173 (S.B.), 11974036 (W.L.), 12222412 (W.L.), 12047503 (W.L.), 12074175 (S.-L.Y.), 11904170 (Z.-Y.D.), and 12004191 (W.W.), the Natural Science Foundation of Jiangsu Province with Grant Nos.~BK20233001 (J.W.), BK20241251 (S.B.), BK20241250 (Y.S.), BK20190436 (Z.-Y.D.), and BK20200738 (W.W.), the Natural Science Foundation of the Higher Education Institutions of Jiangsu Province with Grant No.~23KJB140012 (S.B.), the Postdoctoral Fellowship Program of CPSF under Grant No.~BX20240161 (Y.S.), the China Postdoctoral Science Foundation with Grant No.~2024M751367 (Y.S.), the Jiangsu Funding Program for Excellent Postdoctoral Talent No.~2024ZB021 (Y.S.), the Xiaomi Young Scholars---Technology Innovation Award (S.B.), and the Fundamental Research Funds for the Central Universities (J.Wang; S.B., Grant Nos.~KG202501 and 14380251). We acknowledge the neutron beam time from ANSTO with Proposal Nos.~P9631 and P13772.\\

\bigskip
\noindent {\bf Author contributions}\\
\noindent J.W. conceived the project. S.B. prepared the sample with assistance
from J.L., B.Z., S.C. and H.X. S.B., S.-i.Y., Y.S. and Z.H. carried out the neutron scattering experiments. S.B. and J.W. analysed the experimental data. Y.G., J.Wang, W.W., Z.-Y.D., S.-L.Y., W.L. and J.-X.L. performed the theoretical analyses. J.W., S.B., Y.G., W.L. and J.-X.L. wrote the paper with inputs from all co-authors.\\

\bigskip
\noindent {\bf Competing Interests}\\
\noindent The authors declare no competing interests.\\

\bigskip
\noindent {\bf Additional information}\\
\noindent Correspondence and request for materials should be addressed to J.W. (jwen@nju.edu.cn), J.-X.L. (jxli@nju.edu.cn) or W.L. (w.li@itp.ac.cn).


%

\end{document}